\documentclass[aps,prd,twocolumn,showpacs,amsmath,amssymb]{revtex4}

\usepackage{graphicx}
\usepackage{dcolumn}
\usepackage{bm}

\begin{document}

\title{Nonlinear electrodynamics in Bianchi spacetimes}

\author{Ricardo Garc\'{\i}a-Salcedo}
\email{rsalcedo@fis.cinvestav.mx}
\affiliation{Departamento de F\'{\i}sica,
Cinvestav--IPN, Apartado Postal 14--740, C.P. 07000, M\'exico,
D.F., MEXICO}
\affiliation{Facultad de Ciencias, U. A. E. Mex., Instituto
Literario 100,
50000, Toluca, M\'exico.}
\author{Nora Bret\'on}
\email{nora@fis.cinvestav.mx}
\affiliation{Departamento de F\'{\i}sica,
Cinvestav--IPN, Apartado Postal 14--740, C.P. 07000, M\'exico,  
D.F., MEXICO}

\date{\today}

\begin{abstract}
We study the effects produced by nonlinear electrodynamics in spacetimes
conformal to Bianchi metrics.  In the presence of Born-Infeld type fields
these models accelerate, expand and isotropize. This effect is compared
with the corresponding to a linear electromagnetic field; it turns out
that for the same geometry, Maxwell field does not favours inflation as
much as Born-Infeld field. The behavior of the nonlinear radiation is
analyzed in terms of the equations of state. The energy conditions are
analized as well, showing that the Born-Infeld field violates the strong
energy condition.
\end{abstract}
\vspace{0.5cm}

\maketitle

\section{Introduction}

Born and Infeld proposed their theory for nonlinear electrodynamics (NLE)
in 1934 \cite{BI} with the aim of removing the point charge singularity in
classical electrodynamics. However, soon it became clear that the
Born-Infeld (BI) theory works as well as a phenomenological theory of
quantum electrodynamics in the first approximation. The effective action
for electrodynamics due to one-loop quantum corrections was calculated by
Heisenberg and Euler \cite{HE}, for the low frequency limit; the
corresponding effective Lagrangian have quadratic terms in the
electromagnetic field invariants like the BI Lagrangian.  Moreover,
actions of the BI type have recently been the subject of wide interest in
relation to the result that the effective action for the open string
ending on D-branes can be written in a BI form. In this respect, BI
cosmological solutions could assist in the understanding of brane dynamics
regarding the universe evolution.

These facts make NLE particularly interesting in general relativity,
specially in cosmological theories, as a simple classical model that
incorporates vacuum polarization processes -a possible influence on the
mechanism of the evolution of the early universe; these processes are
expressed in terms of a kind of electric permittivity and magnetic
permeability of vacuum. From this point of view, the exact solutions of
Einstein-Born-Infeld (EBI) equations are worth to study since they may
indicate the physical relevance of nonlinear effects in strong
gravitational and strong electromagnetic fields at early epochs of the
universe. Moreover, in \cite{VMoniz} it was found that BI cosmological
models can be compatible with the recently measured accelerated expansion
of the universe.

One of the topics of interest is to dilucidate to what extent NLE favours
or not an inflationary scenario.  The cosmological inflationary scenario,
as was proposed by Guth \cite{Guth}, has been much favoured due to its
troubleshooting capabilities. Furthermore, it has recently gained the
support of observational evidence \cite{Bernardis}. The conventional
models use some special scalar fields with associated potential to give
rise inflation. Another proposal to produce inflation is to use theories
with unconventional dependence of the Lagrangian on field derivatives
\cite{Altshuler}. In this paper non-linear electrodynamics will be
considered as the dynamical field which drives inflation.

On the other hand, there is no reason to assume that at the onset of
inflation the universe was as isotropic as it seems to be today. It may be
precisely an epoch of accelerated expansion that isotropized the universe.
The most simple anisotropic models are the Bianchi ones.  Inflation in
Bianchi models have been addressed in the Einstein-Cartan theory and with
a cosmological constant \cite{Arts87}. Inflation with a Brans-Dicke field
in Bianchi V has been studied in \cite{Cervantes}; assisted inflation was
investigated in Bianchi I in \cite{Juanma1} and in Bianchi $VI_{0}$ in
\cite{Juanma2}.

In \cite{Gar-Bre} an exact solution of the coupled Einstein-Born-Infeld
equations for anisotropic cosmological scenario was presented. Later,
other cosmological models have been proposed with nonlinear
electromagnetic sources: in \cite{othercos1} it was obtained a homogeneous
and isotropic nonsingular Friedman-Robertson-Walker (FRW) solution
considering a nonlinear generalization of Maxwell electrodynamics as the
source; in \cite{othercos2} it was investigated a Yang-Mills cosmology
with non-Abelian BI action.
   
Here we present a study of the effects produced by a BI type field in an
anisotropic cosmology, based on the analysis of the behavior of the mean
Hubble parameter and the deceleration as well as the energy conditions.  
It turns out that inflation is produced, and in fact, for certain values
of the BI parameter, the inflation is enough to provide the desired
e-folding number that corresponds to the expansion that solves the
cosmological problem of inflation. The model isotropizes at large times,
in such a way that the entry and exit of inflation arise naturally as a
consequence of dynamics. We also present the study of the equations of
state of the nonlinear radiation.
 
The plan of the paper is as follows: In Sec. II the nonlinear
electrodynamics Lagrangian is introduced. In Sec. III the studied metric
is presented; in Sec. IV it is given the corresponding group structure of
the Killing vectors and it is shown that the studied space is conformal to
the Bianchi spaces I, II, III, VII, VIII, IX, for particular values of the
parameters. In section V the kinematical parameter are analyzed as
functions of the synchronous time; comments are given on the e-folding
number; in Sec. VI the effect produced by a linear electromagnetic field
is analyzed. In Sec VII the pressure and energy density of the nonlinear
electromagnetic matter is studied as well as the energy conditions. In
Sec. VIII final remarks are given.
 

\section{Nonlinear Electrodynamics Theory}
  
The nonlinear electromagnetic theory proposed by Born and Infeld is
self-consistent and satisfies all natural requirements. Its Lagrangian
depends in nonlinear form on the invariants of the field,

\begin{equation}
L_{BI}=- \frac{1}{2} P^{\mu \nu} F_{\mu \nu} +{\cal H} (P,Q),
\label{Lagr}
\end{equation}
where ${\cal H} (P,Q)$ is the so called structural function; this function
can vary from one version of nonlinear electrodynamics to other. $P$ and
$Q$ are the invariants associated with the antisymmetric tensor $P_{\mu
\nu}$ that is the generalization of the electromagnetic tensor $F_{\mu
\nu}$. The structural function is constrained to satisfy some physical
requirements: (i) the correspondence to the linear theory $[ {\cal
H}(P,Q)=P+O(P^2,Q^2)]$; (ii) the parity conservation $[{\cal H}(P,Q)={\cal
H}(P,-Q)]$; (iii) the positive definiteness of the energy density (${\cal
H}_{,P} >0$) and the requirement of the timelike nature of the energy flux
vector ($P{\cal H}_{,P}+Q{\cal H}_{,Q}-{\cal H} \le 0$). Condition (iii)
amounts to the fulfilment of the dominant energy condition (DEC) stated as
\cite{Hawk}: $T_{\mu \nu}u^{\mu}u^{\nu} > 0$ for every time direction
$u_{\mu}, (u^{\mu}u_{\mu} < 0)$.  The energy-momentum tensor is given by:

\begin {equation}
T_{\mu \nu}= - {\cal H}_{,P} P_{\mu \alpha}P^{\alpha}{}_{\nu}
+g_{\mu \nu}[2P{\cal H}_{,P} + Q {\cal H}_{,Q}-{\cal H}]. 
\label{eq:tem}
\end{equation}

The structural function for the BI field is given by
\begin{equation}
{\cal H}= b^2- \sqrt{b^4-2b^2P+Q^2},
\label{eq:HBI}
\end{equation}
where $b$ is the Born-Infeld constant that is the ``absolute or maximum
field"; this parameter (comparable to a coupling constant) distinguishes
different theories. For the problem of the electron the value of the
constant is found to be $b=e/r_0^2= 9.18 \times 10^{15}$ e.s.u., while for
the Heisenberg-Euler theory $b=14 \alpha^2/45m^4$, where $\alpha$ is the
fine structure constant and $m$ is the mass of the electron; in string
theory, $b=2 \pi \alpha'$ where $\alpha'$ is the string tension parameter.
Here we shall consider that $b$ is a parameter that can vary in order to
determine if Lagrangians of the type (\ref{Lagr}) are consistent with an
inflationary behavior with a $b$ tuned for that purpose.  The linear limit
is obtained by taking $b \to \infty$, then ${\cal H} =P$ and $P_{\mu
\nu}=F_{\mu \nu}$, coinciding with the Maxwell electromagnetism.

$P_{\mu \nu}$ and $F_{\mu \nu}$ are related through the material or
constitutive equations:

\begin{equation}
F_{\mu \nu}={\cal H}_{,P} P_{\mu \nu}+ {\cal H}_{,Q} \check{P}_{\mu \nu},
\label{eq:mateqs}
\end{equation}
where $\check{P}_{\mu \nu}=- \frac{1}{2} \epsilon_{\mu \nu \alpha
\beta}{P}^{\alpha \beta}$ is the dual of ${P}_{\mu \nu}$. An extensive
treatment on nonlinear electrodynamics can be found in \cite{Pleban}.

\section{EBI solution in an anisotropic spacetime}

We can think of a universe filled with radiation governed by
non-linear dynamics. In \cite{Gar-Bre} it was presented an exact
solution to the coupled system of EBI equations for an anisotropic
cosmological model without further investigations on kinematical
aspects nor inflation. The spacetime is given by the line element

\begin{equation}
ds^2= \frac{1}{\varphi^2} \{ \frac{dz^2}{h}-\frac{dt^2}{s}+ h dy^2+
s( dx+ M dy)^2 \},
\label{lineel} 
\end{equation}
where $x$, $y$ are ignorable coordinates, $h=\alpha + \epsilon z^2$,
$\epsilon, \alpha$ and $l$ are constants, $s=s(t)$, $\varphi = \varphi(t)$
and $M=2lz$. The motivation to choose a metric of the form (\ref{lineel})
is that it is a type D metric which for the stationary axisymmetric case
(one spacelike Killing vector and one timelike Killing vector) solutions
are known of coupled Einstein-Born-Infeld fields \cite{Pleban0}. We
expected, on the other side, that metric (\ref{lineel}) be related to
$G_2$ cosmologies, however the analysis of the Killing vectors shows (see
next Section) that metric (\ref{lineel}) possesses four Killing vectors
and that it is conformal to Bianchi metrics. The spacetime (\ref{lineel})
is coupled to an NLE energy-momentum tensor given by Eqs. (\ref{eq:tem})
and (\ref{eq:HBI}), such that the two nonvanishing components of the
electromagnetic field have been aligned with the two double repeated
directions of the Weyl tensor and parametrized by $P_{12}= i \breve{H},
P_{34}=D$, where $\breve{H}$ is the magnetic field and $D$ is the electric
displacement; the invariants of the BI field are then given by
$P=-\frac{1}{2}(D^2-\breve{H}^2)$ and $Q=-i \breve{H} D$. We introduce the
function $\nu(t)$ defined by $e^{\nu}= \sqrt{b^2- \breve{H}^2}/
\sqrt{b^2+D^2}$.

The solution to the coupled EBI equations
is given by the metric function $s(t)$ in terms of $\varphi (t)$
(see \cite{Gar-Bre} for details),

\begin{equation}
s(t)= \varphi^3 \dot{\varphi} \{ c_1 - 2b^2 \int{ \frac{( e^{\nu}
-1)}{\varphi^4 \dot{\varphi}^2}dt} - \epsilon \int{ \frac{dt}{\varphi^2
\dot{\varphi}^2}} \},
\label{ssol}
\end{equation}
where dot simbolizes derivative with respect to $t$, $c_1$ is an
integration constant related to the solution in vacuum, the second term
corresponds to the BI contribution and the last term is related to the
spatial curvature, being $ \epsilon = 1$ for a closed universe, $ \epsilon
=0$ for spatially flat universe and $ \epsilon = -1$ for an open universe.  
Moreover, $\varphi (t)$ and $\nu(t)$ must satisfy the equations
$\ddot{\varphi}+l^2 \varphi =0$ and $\dot{\nu} \varphi-2 \dot{\varphi}
(1-e^{-2 \nu}) =0$.
  
The system of EBI equations admits two possible solutions: $e^{\nu}=
\sqrt{1 \pm \varphi^4}$. The case $e^{\nu}= \sqrt{1 - \varphi^4}$ is the
one that admits a cosmological interpretation, since in this case $s(t) >
0$. The case $e^{\nu}= \sqrt{1 + \varphi^4}$ corresponds to a metric
function $s(t) < 0$ which changes the signature of the metric
(\ref{lineel}) to a stationary axisymmetric one. In this work we restrict
to the range of $t$ for which $s(t) > 0$; otherwise the signature of the
metric changes and another interpretation must be considered.
 
The condition that $e^{\nu}$ be real imposes the restriction that
$\varphi ^4 < 1$; in terms of the fields it amounts to $\breve{H} <
b$ and $D < b$, that is, the fields must not reach the maximum field
strenght $b$. Also this condition determines the range of the
coordinate $t$ with $ 0 \le \varphi (t) <1$.

Moreover, the only nonvanishing Weyl scalar is given by

\begin{equation}
\psi_2= \frac{\varphi^2}{12} [\ddot{s}-8sl^2-2 \varepsilon +6il \dot{s}].
\end{equation}

The invariants of this spacetime are $I=3 \psi_{2}^2$, $J=- \psi_2^3$,
in such a manner that if $\psi_2$ diverges, so do the invariants. For this
solution, $\psi_2$ is well behaved and does not present divergences.
Up to here these results were derived in \cite{Gar-Bre} and what follows
is the contribution of the present paper.

\section{Characterization of the spacetime}
 
In this section it is shown that the metric (\ref{lineel}) is
conformal to Bianchi spaces I, II, III, VII, VIII and  IX.
 
The metric
(\ref{lineel}) possesses four Killing vectors:

\begin{eqnarray}
\xi_1&=& [0,1,0,0], \qquad \xi_2=[0,0,1,0],\nonumber\\ \xi_3&=&[\sqrt{h}
S{(\sqrt{\epsilon \alpha}y)}, -\frac{2 l a C{(\sqrt{\epsilon
\alpha}y)}}{\sqrt{h \epsilon \alpha}}, -\frac{\epsilon z C{(\sqrt{\epsilon
\alpha}y)}}{\sqrt{h \epsilon \alpha}},0], \nonumber\\ \xi_4&=&[\sqrt{h}
C{(\sqrt{\epsilon \alpha}y)}, -\frac{2 l a S{(\sqrt{\epsilon
\alpha}y)}}{\sqrt{h \epsilon \alpha}}, -\frac{\epsilon z S{(\sqrt{\epsilon
\alpha}y)}}{\sqrt{h \epsilon \alpha}},0],
\end{eqnarray} 

where $S$ stands for $\sinh$ and $C$ for $\cosh$.  This set corresponds to
a four-generator group with invariance subgroups of spatial rotations or
boosts. From the above expressions is easy to show the commutation rules,
$[\xi_1,\xi_3]=\sqrt{\epsilon \alpha} \xi_4 ; \quad
[\xi_1,\xi_4]=\sqrt{\epsilon \alpha}\xi_3; \quad
[\xi_3,\xi_4]=\sqrt{\epsilon \alpha} \xi_1.$

The metric (\ref{lineel}) is conformal to several Bianchi spaces, it
is demonstrated by performing the transformation

\begin{equation}
\frac{dz^2}{h(z)}=dy^2, \quad \frac{dt^2}{s(t)}=dT^2, \quad y=z,
\quad x=x,
\end{equation}
with this transformation the metric, Eq. (\ref{lineel}), acquires the form

\begin{equation} 
\varphi^2 ds^2=- dT^2 + s(T)(\sigma^1)^2 + dy^2 +h(y) dz^2,
\end{equation}  
where we named $dx+M(y)dz=\sigma^1$. Bianchi types IX, II, III and VIII
can be identified according to  Table \ref{table1} (cf. Eq. (11.4) in
\cite{Kramer}).


\begin{table}
\caption{\label{table1}\small{Values of $\alpha$ and $\epsilon$ 
for which the studied metric is conformal
to the Bianchi types}}
\begin{ruledtabular}
\begin{tabular}{|c|c|c|c|c|}
{$\sigma^1$} &{$h(y)$} & $\alpha$ & $\epsilon$ & Bianchi type\\ \hline
\hline
{$dx+ 2l\cos{y}dz$}& { $\sin^2{y}$} &  1 & -1 &  G$_3$IX \\
{$dx+ 2l y dz$} & 1 & 1 & 0 & G$_3$II \\
{$dx+ 2l\cosh{y}dz$} & { $\sinh^2{y}$} & -1 & 1 & G$_3$III, G$_3$VIII \\
\end{tabular}
\end{ruledtabular}
\end{table}

Furthermore, in the case $M=0, (l=0)$ the line element (\ref{lineel})
becomes conformal to G$_3$I or G$_3$VII for $\epsilon=0$ and G$_3$III for
$\epsilon=1$.

Also, with a coordinate transformation, the metric (\ref{lineel}) can
be identified as a hypersurface-homogeneous spacetime with a four
dimensional group of isometries, $G_4$, with three dimensional subgroups
that can be of spatial rotations or boosts ($S_3$ or $T_3$),

\begin{equation}
\varphi ^2 ds^2 = -\frac{4 d \zeta d \bar{\zeta}}{(1+\epsilon \zeta
\bar{\zeta})^2} - \frac{dt^2}{s(t)}+s(t) \left( dx +2il \frac{\zeta
d \bar{\zeta}-\bar{\zeta} d\zeta}{(1+\epsilon \zeta \bar{\zeta})}
\right) ^2,
\label{11.11}
\end{equation}

The needed transformation to pass from Eq. (\ref{lineel}) to
Eq.(\ref{11.11}) is (t remains the same):

\begin{equation}
z=-a \frac{(1- \epsilon \zeta \bar{\zeta})}{(1+ \epsilon \zeta
\bar{\zeta})}, \quad y=\frac{i}{2a \epsilon} \ln
\frac{\bar{\zeta}}{\zeta}, \quad x= x+\frac{il}{\epsilon} \ln
\frac{\bar{\zeta}}{\zeta}\, .
\end{equation}

In the form (\ref{11.11}) it is manifest the existence of a
two-dimensional submanifold of constant curvature.


\section{Inflation produced by the BI field}
      
In this section we study the kinematical behavior in the asymptoyic
regimes of $\varphi \to 1$ and $\varphi \to 0$, that correspond to $t \to
0$ and $t \to \pi/2$, respectively; in a synchronous reference system with
synchronous time given by $dT= \frac{dt}{\varphi \sqrt{s}}$, $\varphi \to
1$ and $\varphi \to 0$ correspond to $T \to 0$ and $T \to \infty$,
respectively. In this analysis we shall restrict to the effect produced by
the BI field, meaning that in the metric function $s(t)$, Eq.
(\ref{ssol}), we choose $c_1=0= \epsilon$, then

\begin{equation}
s(t)= - 2b^2 \varphi^3 \dot{\varphi} \int{ \frac{(e^{\nu}
-1)}{\varphi^4 \dot{\varphi}^2}dt} =b^2 S_{BI}
\label{ssolBI}
\end{equation}

Moreover, we approximate $e^{\nu}-1 = \sqrt{1- \varphi^4}-1 \approx
- \varphi^4 (1+ \varphi^4/4 + \varphi^8/8)/2$, approximation valid since
$\varphi^4 <1$. Then in the asymptotic regimes the limits for the
metric function $s(t)$ and its derivative are

\begin{eqnarray} 
s(t \to 0)  \to 11b^2/8;&& \qquad \dot{s}(t \to 0) \to 0. \nonumber\\
s(t \to \pi/2)  \to 0;&& \qquad \dot{s}(t \to \pi/2) \to 0.
\label{limits}
\end{eqnarray}
 
We now proceed to analyze the Hubble, deceleration and shear parameters.  
In the synchronous reference system where $g_{00}=1$ and $g_{0i}=0$, we
introduce a spatial metric tensor $h_{ab}$. The volume element in three
space is $\sqrt{h}=\varphi^{-3}\sqrt{s}$, ($h=$ det $h_{ab}$), while the
volume expansion or mean Hubble parameter is given by

\begin{equation}
H= \frac{\varphi \sqrt{s}}{6}
(\frac{\dot{s}}{s}-6\frac{\dot{\varphi}}{\varphi}),
\label{Hubble}
\end{equation}
The Hubble parameter has been calculated using the timelike unit vector,
$U^{\alpha}= \varphi \sqrt{s(t)} \delta^{\alpha}_{t}$, or in the null
tetrad formalism, $U^1=U^2=0$ and $U^3=-U^4=\frac{1}{\sqrt{2}}$. In Eq.
(\ref{Hubble}) $H$ is a function of $t$. The role of the constant $c_1$ in
the metric function $s(t)$, Eq.(\ref{ssol}), when substituted in
(\ref{Hubble}) is to shift the time for which the expansion becomes null.
The instant when $H=0$ can be placed before, at or after $t=0$ by changing
the value of $c_1$.
   
\begin{figure}
\includegraphics[width=8cm,height=6cm]{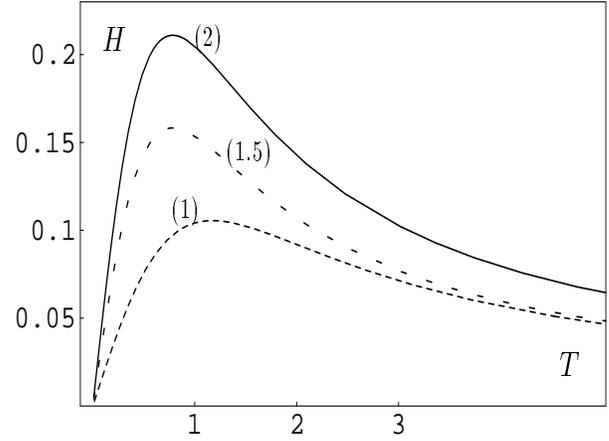}
\caption{\label{fig1}
It is shown the mean Hubble parameter $H$ as function of the synchronous
time $T$ for distinct values of the BI parameter $b$ (shown in
parenthesis). It can be observed that the expansion has a maximum and
after some time decays to a constant value. In this plot $\varphi=.8
\cos{t}, \epsilon=0, c_1=0.$
} 
\end{figure}

In Fig.  \ref{fig1} it is shown a parametric plot of the mean Hubble
parameter $H(T)$ in terms of the synchronous time. $H(T)$ is plotted for
several values of the Born-Infeld parameter $b$. It is clear that as
greater is $b$, more expansion is produced. After some time $H(T)$ goes to
a constant.  To generate the graphics we have chosen $\varphi = A \cos(lt
+ \beta)$ as the solution of $\ddot \varphi +l^2 \varphi =0$ and we have
put the phase to zero.

The expression of the Hubble parameter considering only the
BI contribution clearly shows its $b$ dependence,

\begin{equation}
H_{BI}= b \frac{\varphi  
\sqrt{S_{BI}}}{6}
(\frac{\dot{S_{BI}}}{S_{BI}}-6\frac{\dot{\varphi}}{\varphi}).
\end{equation}

The analysis of $H_{BI}$ in the asymptotic regimes shows that
both at $t \to 0$ ($T \to 0$) and at $t \to \pi/2$ ($T \to \infty$)
$H_{BI} \to 0$, so the expansion occurs during a limited period of time.
  
The behavior of the Hubble parameter is not enough to determine whether
the model inflates or not, rather we must determine the sign of the
deceleration parameter, and it must be negative if inflation is going to
happen. The deceleration is $q=- \theta^{-2}(3 \theta_{\alpha}U^{\alpha} +
\theta^{-2})$, where $\theta$ is the expansion of a timelike vector; in
terms of $s(t)$ and $\varphi (t)$ the deceleration is given by

\begin{equation}
q=3 \theta^{-2}(\dot{\varphi} \varphi \dot{s} + 
\varphi^2(\dot{s}^2/4s- \ddot{s}/2-3l^2s))-1.
\label{q} 
\end{equation}
  
Using the field equation $2b^2(e^{- \nu}-1)= \varphi^2
\ddot{s}/2-2\dot{\varphi} \varphi \dot{s}+3s(\dot{\varphi}^2+l^2 \varphi^2
)$ and approximating $e^{-\nu} \approx 1+ \varphi^4/2 +3 \varphi^8/8$,
which is valid since $\varphi^4 < 1$ and considering only the BI effect we
obtain

\begin{equation}   
q= \frac{b^2}{\theta^{2}} \{ \frac{\varphi^2 \dot{S_{BI}}^2}{2S_{BI}} -
3 \varphi^4 (1+ \frac{3}{4} \varphi^4+ \frac{5}{8} \varphi^8) \}.
\label{qBI}
\end{equation}
   
\begin{figure}
\includegraphics[width=8cm,height=6cm]{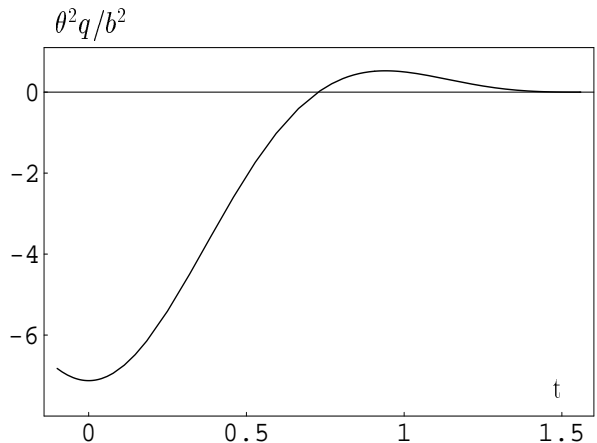}
\caption{\label{fig2}
The qualitative behavior of the deceleration parameter $q$ is displayed in
this figure: at the beginning there is acceleration followed by a
deceleration period, $q >0$, and finally, as $t \to \pi/2$, $q \to 0$. The
plot corresponds to $q \theta^2/b^2$ and $s(t)= b^2 S_{BI}$, with
$\varphi=\cos{t}$.
}
\end{figure}

The sign of the deceleration parameter is determined by the term in curly
brackets, it is the second term, which can make $q$ to be negative, since
$S_{BI}, \varphi^2$ and $\dot{S_{BI}}^2$ are positive.  It is also
illustrative that the asymptotic behavior of $q$ as $t$ tends to zero is
$q(t \to 0) \to - \infty$; the infinite is due to the fact that as $t$
tends to zero the expansion $\theta$ also tends to vanish; generically for
the BI contribution we can say that $q$ is negative as the time tends to
zero. Regarding the asymptotic regime of $t \to \pi/2$ ($T \to \infty$),
$q$ tends to vanish, $q(t \to \pi/2) \to 0$, with this showing that the
deceleration stops as the synchronous time goes to infinity. Fig.
\ref{fig2} shows the behavior of $q$, the plot corresponds to $q \theta^2
/b^2$.
  
\begin{figure}
\includegraphics[width=8cm,height=6cm]{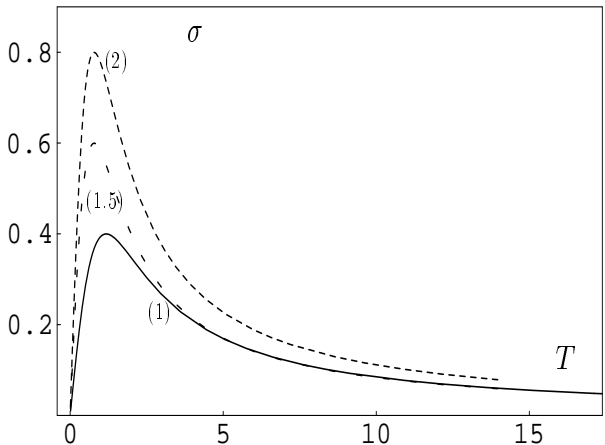}
\caption{\label{fig3}
 It is shown a parametric plot of the shear $\sigma$ and the synchronous
time $T$, for several values of $b$ (shown in parenthesis). The model 
isotropizes after some time and the maximum of the shear is higher for
larger values of the parameter $b$. Here $\varphi =.8 \cos{t},
c_1=0$ and $\epsilon=0$.}
\end{figure}
  
The behavior described above of the expansion and deceleration parameters
is qualitatively the same for the cases $\epsilon = 1, 0, -1$ (different
Bianchi spaces). For distinct values of $\epsilon$ the inflationary period
changes its duration: for $\epsilon =- 1$, $q$ is negative but greater in
absolute value during a shorter period. For $\epsilon = 1$, $q$ is
negative but smaller in absolute value than for $\epsilon =-1$, but $q$ is
negative during a longer period. The point where $q$ diverges to negative
infinity is the same for which $H=0$, then in this respect the shifting in
time also applies for $q$: by changing the value of $c_1$ in the function
$s(t)$, the value of $t$ for which $q$ diverges is shifted.

It has been shown that there exists an accelerated expansion produced by
the BI type field. How much does inflate a spacetime is specified by the
e-folding number $N$. The cosmological problem of expansion is solved with
a minimum e-folding number of about 70, that amounts to an expansion of
$10^{30}$. The number of e-folds of growth in the scale factor that occurs
from $t_1$ to $t_2$ is given by

\begin{equation}
N= \int_{t_1}^{t_2} H_{BI}dt= b \int_{t_1}^{t_2} { \frac{\varphi
\sqrt{S_{BI}}}{6}
(\frac{\dot{S_{BI}}}{S_{BI}}-6\frac{\dot{\varphi}}{\varphi})}dt,
\end{equation}
Since the integrand is positive the number $N \sim 70$ can be reached by
adjusting the parameter $b$; for $b \sim 10^3$ it is obtained the desired
$N$.

\begin{figure}
\includegraphics[width=8cm,height=6cm]{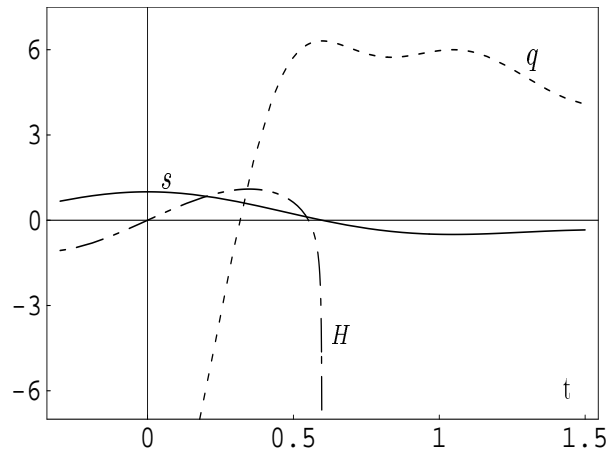}
\caption{\label{fig4}
The expansion $H$, deceleration $q$ and the metric function $s(t)$
produced by the linear electromagnetic field are shown in this figure. In
this case inflation occurs during a short period of time which ends when
the deceleration becomes positive (at $t=0.32$).  Here $\varphi = \cos{t},
C=1, c_1=0$ and $\epsilon=0$. }
\end{figure}

In relation with the isotropy of the model, it has been shown within the
framework of general relativity \cite{Col-Haw} that if the dominant energy
conditions and positive pressure criterion are satisfied then the universe
will isotropize starting from Bianchi I, V, VII$_o$, and VII$_h$ models.
The anisotropic character is given by the shear, $\sigma$, then the
necessary condition for eventualy obtain a FRW cosmology is the vanishing
of the shear. For our model the shear is given by $\sigma^2= \varphi^2
\dot{s}^2/6s$. Then a requirement for the model to isotropize is that the
derivative of $s(t)$ goes to zero after some time.  The tendency of the
shear at the asymptotic regimes is such that it vanishes both as $t$ goes
to zero and as $t$ approximates $\pi/2$, $\sigma (t \to 0) \to 0$ and
$\sigma (t \to \pi/2) \to 0$. Then the shear grows at early times in the
model and as $t \to \pi/2$ ($T \to \infty$ in the synchronous time) it
gradually decays reaching zero as the synchronous time goes to infinity.
In Fig. \ref{fig3} the shear is shown in terms of the synchronous time
$T$, for several values of the BI field;  the peak of $\sigma$ is greater
as larger is the parameter $b$. This plot shows how these BI cosmologies
isotropizes. Only the BI effect is considered in the graphics, meaning by
this that the constants in $s(t)$, Eq (\ref{ssol}), are $c_1=0= \epsilon$.

\section{Inflation produced by a
linear electromagnetic field}
      
It is interesting to compare the effects produced by the BI field with
those corresponding to the limit of linear electromagnetic field.  This
effect is obtained by taking $b \to \infty$. The process of taking this
limit involves at the same time to take it in the function $e^{\nu}=
\sqrt{b^2- \breve{H}^2}/ \sqrt{b^2+D^2}$, in such a manner that when $b
\to \infty$ we have

\begin{eqnarray}
2b^2(e^{- \nu}-1)&& \to D^2+ \breve{H}^2, \nonumber\\
-2b^2(e^{\nu}-1)&& \to  D^2+ \breve{H}^2.
\end{eqnarray}
 
Using this limits in the expressions for the metric function $s(t)$ as
well as in the expansion and deceleration we obtain the corresponding
quantities when a linear electromagnetic field is acting on the metric
(\ref{lineel}). They are:

\begin{eqnarray}
s(t)&=&\varphi^3 \dot{\varphi} \{ c_1 + C^2 \int{
\frac{dt}{\varphi^4 \dot{\varphi}^2}} - \epsilon \int{
\frac{dt}{\varphi^2 \dot{\varphi}^2}} \},
\label{sMax} \\
q&=& \frac{2\dot{s}^2 \varphi^2-12 s
C^2}{(6s\dot{\varphi}-\dot{s}{\varphi})^2},
\end{eqnarray} 
where $C^2=D^2+ \breve{H}^2$ is related with the energy density of the
linear electromagnetic field; the expansion $H$ is given as in Eq.
(\ref{Hubble}) with $s(t)$ from Eq. (\ref{sMax}).  
  
Under the influence of the linear electromagnetic field, the model can
present a positive expansion and negative deceleration simultaneously,
then producing inflation, however, the period of inflation is shorter than
the corresponding to the BI field. Morever, the interpretation of the
spacetime is not clearly cosmological for the periods when $s < 0$
($t>0.5$ in Fig. \ref{fig4}).  The conclusion is that, for the same
geometry, the linear field does not favour inflation as much as BI does.
The effect is shown in Fig. \ref{fig4}.


\begin{figure}
\includegraphics[width=8cm,height=6cm]{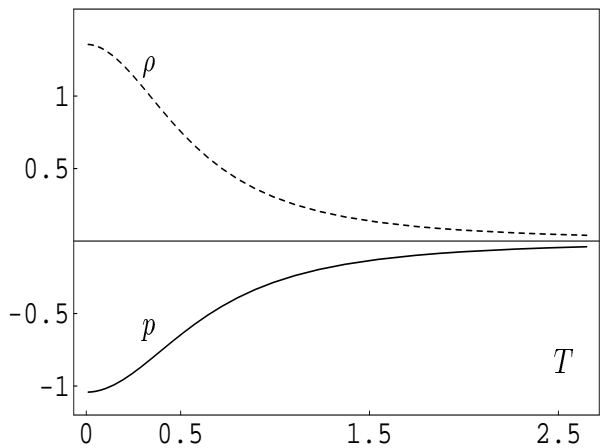}
\caption{\label{fig5}
In this plot it is shown the two distinct eigenvalues of $T^{ab}$:
$\rho=p_1=T_{34}$ and $p= p_2=p_3=T_{12}$. $\rho$ is the energy density and
$p_1, p_2, p_3$ are the pressures in the three spatial directions that
measures an observer at the point $p^*$, whose world line has as tangent
the timelike vector $E_4$. The observer measures in one direction a stiff
fluid of positive pressure ($p=\rho$) and in the other two directions there
is a kind of ``vacuum" state with negative pressures ($p_2=p_3 <0$).
}  
\end{figure}

\begin{figure}
\includegraphics[width=8cm,height=6cm]{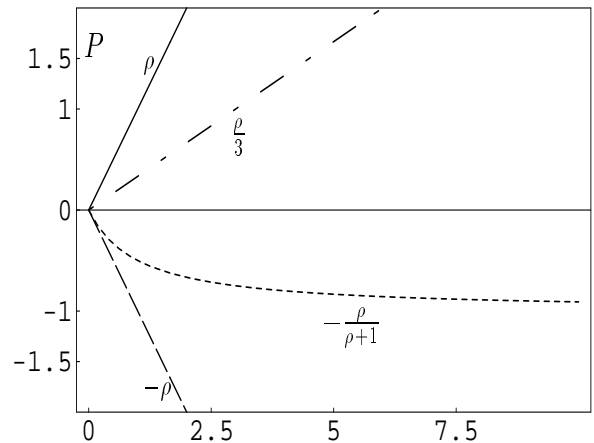}
\caption{\label{fig6}
It is shown the behavior of the matter with the pressure $p$ as function of
the energy density $\rho$. Both pressure and energy density are scaled as
$p/b^2$ and $\rho /b^2$. Also are shown, for comparison, the ``vacuum-like"
state of $p=- \rho$ as well as the equation of state of the radiation, $p=
\rho/3$. This behavior occurs during the period that the spacetime is
anisotropic.
}
\end{figure}

\section{Equation of state and energy conditions}
  
We can ask about the behavior of the nonlinear radiation or the
relationship between the energy density and the pressure. Such relation
can not be the equation of state of a perfect fluid because the algebraic
structure of the energy momentum tensor $T_{\mu \nu}$ in our problem is
not compatible with that of a perfect fluid. To correspond to a perfect
fluid $T_{\mu \nu}$ should have three repeated eigenvalues. Rather, in the
studied case, the energy momentum tensor possesses two double repeated
eigenvalues that for an anisotropic space they can be interpreted as a
different behavior of matter depending on the direction. The two double
repeated eigenvalues of $T_{\mu \nu}$ are
   
\begin{equation}
T_{12}= 2b^2(e^{ \nu}-1), \quad T_{34}= 2b^2(e^{- \nu}-1).
\end{equation}
    
When the energy-momentum tensor has only one timelike eigenvector ($E_4$)
one may express the components $T^{ab}$of the energy-momentum tensor at a
point $p^*$ with respect to an orthonormal basis in the canonical form
$T^{ab}=$ diag[$p_1,p_2,p_3, \rho$], where $\rho$ represents the energy
density measured by an observer whose world line at $p^{*}$ has as unit
tangent vector $E_4$; the eigenvalues $p_{\alpha}$ represent the pressures
in the three spatial directions.  The corresponding eigenvalues are $\rho
= p_{1}=T_{34}$ and $p_2=p_3=T_{12}$, they are plotted in Fig. \ref{fig5}.
One of the eigenvalues corresponds to the energy density and the pressure
in one direction, while the other one corresponds to the pressure in the
other two directions.  The observer at $p^*$ measures in one direction a
stiff matter $\rho = p_1 >0$ while in the other two directions the
pressures are negative $p_2=p_3 =p < 0$. Fig \ref{fig5} also shows that as
the time $T$ grows, the eigenvalues tend to the same value but with
negative signs, i. e. $ \rho +3p=0$, it implies that the trace of the
energy-momentum tensor becomes zero, and that the matter has the equation
of state $p=- \frac{\rho }{3}$. Note that it coincides with the equation
of state of the gas of Nambu-Goto strings in three spatial dimensions.

The energy density is given by

\begin{equation}
\rho= \frac{2b^2}{8 \pi} \{ \sqrt{\frac{b^2+D^2}{b^2- \breve{H}^2}} -1 \},
\label{energyden}
\end{equation}
 
Eq. (\ref{energyden}) shows that the growth of $\rho$ with the fields
$\breve{H},D$ is slower than that in the Maxwell theory, where $\rho \sim
\breve{H}^2+ D^2$. While the negative pressures are:

\begin{equation}
p_2=p_3=p=\frac{2b^2}{8 \pi} \{ \sqrt{\frac{b^2- \breve{H}^2}{b^2+D^2}} -1
\},
\label{presion}
\end{equation}
or in terms of $\rho$, for these two directions and scaling $\rho /b^2$
and $p/b^2$, we have

\begin{equation}
p= - \frac{\rho}{ \rho +1},
\label{eq.state}
\end{equation}
   
The behavior of $p( \rho)$ is illustrated in Fig. \ref{fig6}. In these two
directions the matter behaves as a ``vacuum-like" state.  The observer at
$p^*$ sees that there is a coexistence of an expanding-contracting
asymmetry: for $p= - \rho/(\rho +1)$ produces expansion in two directions
(negative pressures) whereas contraction prevails in the third direction,
$p= \rho$, producing together a global expansion shown in Fig.1.
  
In the two directions of negative pressure, the compressibility is
negative, $dp/d \rho <0$; it is well known that negative compressibility
leads to a strongly inhomogeneous mixture of different phases. From Eqs.
(\ref{energyden}) and (\ref{presion}), one has also that

\begin{equation}
\rho= -2b^2 p/(8 \pi p + 2 b^2),
\end{equation} 
 
This equation indicates that for values of the pressure such that $p= -
\frac{2b^2}{8 \pi}$ there occurs a critical point in the energy density,
$\rho$ diverges for this value of the pressure. It corresponds to the
values of the fields that superate the maximum field strenght, $\breve{H}
\ge b$. The instability in the energy density indicates a possible phase
transition at $\rho > \rho_{crit}$. It is known that Born-Infeld theory
for values of the invariants much greater that the limit field, $P >>
b^2$, i.e. in the ``super-quantum limit", is equivalent to the theory of
magnetic strings. Hence in spacetime regions of strong field the
self-trapping of the field into condensate of closed Nambu strings should
be expected \cite{Tze}.
  
In \cite{Altshuler} Altshuler investigated nonlinear electrodynamics as an
alternative way to inflation and the possibility of anti-inflation.
Altshuler addressed the case of the Lagrangian depending in nonlinear way
only on one of the invariants of the electromagnetic field; however some of
the conclusions driften there applies to our case for the directions of
observed negative pressures. One of the conclusions of his study is that
for sufficiently high density the radiation, governed by nonlinear
dynamics, has a tendency to condensation when the effective pressure is
negative. This condensate state is the responsible for inflation. However,
strictly speaking, to obtain an effective equation of state we should make
a statistical approach, like a kind of average in the three directions.

\subsection{Energy Conditions}

In this subsection the energy conditions for the model are analyzed.

The dominant energy condition (DEC) states that {\it for every timelike
vector, $W_a$, $T^{ab}W_{a}W_{b} \ge 0$ and $T^{ab}W_{a}$ is a nonspacelike
vector} \cite{Hawk}. This condition means that to any observer the local
energy density appears nonnegative and the local energy flow vector is
non-spacelike. In the case of the Born-Infeld energy momentum tensor DEC is
a requirement of a physically reasonably nonlinear structural function and
its fulfilment is guaranteed by condition (iii) in Sec. II. For the
timelike unit vector $U^{\alpha}$ DEC amounts to

\begin{equation}
T_{ab}U^a U^b=2T_{34}U^3U^4= 2 b^2 (e^{-\nu}-1) \ge 0,
\label{DEC}
\end{equation}
clearly DEC is fulfilled if $e^{\nu}=\sqrt{1-\varphi^4} < 1$, which in
fact is the case we addressed in this work ($\varphi^4 < 1$).
 
The strong energy condition (SEC), as stated in \cite{Seno}, says: {\it A
spacetime satisfies the strong energy condition if $R_{ab}u^{a}u^{b} \ge
0$, for all causal vectors $u^a$}. The same condition is called by Hawking
as the {\it timelike convergence condition} \cite{Hawk}.

In the studied case the traceless Ricci tensor $R_{\mu \nu}$ has two
double repeated eigenvectors such that in the Segr\`e classification
$R_{\mu \nu}$ corresponds to A1[(11)(1,1)].  Actually the two nonvanishing
components of the traceless Ricci tensor are

\begin{equation}
R_{12}=2b^2(1-e^{\nu}), \quad R_{34}=2b^2(1-e^{- \nu}). 
\end{equation}
      
For the BI field SEC amounts to $R_{ab}U^{a}U^{b}=R_{34}=2b^2(1- e^{-
\nu}) < 0$, which clearly violates SEC and the reason is the existence of
the negative pressures ($p_2=p_3 <0$).  This fact must not disturb us,
since inflationary models driven by scalar inflaton fields violate SEC
during the inflationary epoch. The minimally coupled scalar field violates
SEC and indeed curvature-coupled scalar field theory also violates SEC.
This point has been discussed recently in \cite{Seno} and \cite{Visser}.


\section{Final Remarks}
    
In this paper we showed that the Born-Infeld nonlinear electromagnetic
field induces inflation in anisotropic cosmological models that are
conformal to several Bianchi spaces.  Moreover, the e-folding number to
obtain enough inflation to solve the horizon and flatness problems can be
provided by nonlinear electrodynamics with a fine tunning of the parameter
$b$.  Note that changing the value of $b$ imply physically different
nonlinear electrodynamic theories coupled to gravity, since the coupling
constant is different. The asymptotic behavior of the kinematical
parameters shows that the inflation occurs in a limited period of time and
that the model isotropizes, since as the synchronous time goes to infinity
the expansion, deceleration and shear tend to vanish.

In contrast, the linear electromagnetic field inflates the model in a very
restricted period of time and the range of the coordinate $t$ valid for a
cosmological interpretation is also reduced.
   
The analysis of the nonlinear electromagnetic radiation in the anisotropic
space exhibites expansion in two directions, associated to negative
pressures that could originate condensated states which allow inflation.
In the third direction the equation of state corresponds to stiff matter,
this fact could indicate compression in such direction.

We also analyzed the energy conditions and it turned out that the
strong energy condition SEC is violated by the BI field.  The
violation of this condition could has as a consequence the absence of
an initial singularity, since the fulfiling of SEC is related with   
the convergence of neighbouring geodesics. However, in \cite{Borde} 
it was obtained the result that inflationary spacetimes are not past
complete, i.e. the corresponding geodesics are incomplete and then a
singularity exists.  Therefore we wonder if this spacetime is   
singular. It is work in progress if in the case we addessed the 
nonlinear electromagnetic field does prevent the existence of an
initial singularity.

Being the nonlinear electromagnetic field consistent with an inflationary
scenario, it might be that the BI field or another field with a Lagrangian
of this type was present at early epochs of the universe, probably not
alone, but accompanied with some scalar fields and that together, in a
kind of ``assisted inflation", cooperate to produce the appropriate
initial conditions for a FRW-type universe. However to address the
standard picture of inflation the question arises on How can we consider
quantum fluctuations of the BI field during inflation?, since in fact BI
theory corresponds to an effective modification induced by quantum
fluctuations. Also we must investigate if within the BI theory is it
possible to have reheating in the universe; this lead us to the question
of creation of particles and quantization of the BI field; this study
was initiated by Born and Infeld themselves in 1935 and they showed that
the commutation rules for the field components were very similar to the
ones of quantum mechanics. Alternatively, one could adopt a squeme like
the proposed in higher-derivative gravity theories and to look for a
conformal equivalence between BI field coupled to gravity on one side and
the Einstein theory with a scalar field on the other side, and then
proceed as in the standard inflationary treatments. All these are
speculations that have to be worked out.


\begin{acknowledgments}
The authors acknowledge one of the referees for a careful reading of our
manuscript and for valuable suggestions. R. G-S. gratefully acknowledges
fruitful discussions with J. M. M. Senovilla. R. G-S. was partially
supported by FC-UAEMex. This work was partially supported by CONACyT
(M\'exico) under project 40888A-1.

\end{acknowledgments}


\begin{thebibliography}{99}  

\bibitem{BI}{M. Born and L. Infeld, Proc. Royal Soc. London
{\bf A144}, (1934) 425}

\bibitem{HE}{W. Heisenberg and H. Euler,Z. Phys. {\bf 98} (1936) 714.}

\bibitem{VMoniz} P. Vargas Moniz, Phys. Rev. {\bf D66}, (2002) 103501;
Class. Quant. Grav. {\bf 19} (2002) L127-L134.

\bibitem{Guth}{A. Guth, Phys. Rev. {\bf D23} (1981) 347.}

\bibitem{Bernardis}{P. de Bernardis {\it et al}, Nature {\bf 404} (2000)
955.}

\bibitem{Altshuler}{B. L. Altshuler, Class. and Quantum Grav. {\bf 7}
(1990) 189.}

\bibitem{Arts87}{L. Jensen and J. Stein-Schabes, Phys. Rev. {\bf D35}
(1987) 1146;
M. Demianski, R de Ritis, G. Platania, P. Scudellaro and C.
Stornaiolo, Phys. Rev. {\bf D35} (1987) 1181.}
 
\bibitem{Cervantes}{J. L. Cervantes-Cota, Class. Quant. Grav.
{\bf 16} (1999) 3903.}

\bibitem{Juanma1}{J. M. Aguirregabiria, A. Chamorro, L. P. Chimento,
and N. Zuccal\'a, Phys. Rev. {\bf D62} (2000) 084029.}


\bibitem{Juanma2}{J. M. Aguirregabiria, P. Labraga and R. Lazkoz, 
Gen. Rel. Grav. {\bf 34} (2002) 341.}


\bibitem{Gar-Bre}{R. Garc\'{\i}a-Salcedo and N. Bret\'on,
Int. J. Mod. Phys., {\bf A15}, (2000), 4341.}

\bibitem{othercos1}{V. A. De Lorenci, R. Klippert, M. Novello, J. M.
Salim,  Phys. Rev. {\bf D65}, (2002) 063501.} 

\bibitem{othercos2}{V. V. Dyadichev, D. V. Gal\`tsov, A. G. Zorin, M. Yu.
Zotov, Phys. Rev. {\bf D65}, (2002) 084007.}
 
\bibitem{Hawk}{S. W. Hawking and G. F. R. Ellis, {\it The large scale
structure of spacetime}  (Cambridge Univ. Press, 1973.)}

\bibitem{Pleban}{J. F. Plebanski, {\it Lectures of Nonlinear
Electrodynamics}
monograph of the Niels Bohr Institute Nordita, Copenhagen (1968).}

\bibitem{Pleban0}{A. Garc\'{\i}a, H. Salazar and J. F. Pleba\~nski,
Nuovo Cim. {\bf 84} (1984) 65.}
 
\bibitem{Kramer}{D. Kramer, H. Stephani, M. A. H. MacCallum and E. Herlt,
{\it
Exact Solutions of Einstein's Field Equations}, (Cambridge University
Press, Cambridge, 1980.)}

\bibitem{Col-Haw}{C. B. Collins and S. W. Hawking, Astrophys. J. {\bf 180}
(1973),317.}

\bibitem{Tze} {H. C. Tze, Nuovo Cimento {\bf A22} (1974) 507}

\bibitem{Seno}{J. M. M. Senovilla, Gen. Rel. Grav., {\bf 30}
(1998) 701-848}

\bibitem{Visser}{M. Visser and C. Barcel\'o, {\it Energy conditions and
their cosmological implications}, gr-qc/0001099.}

\bibitem{Borde} {A. Borde, A. H. Guth and A. Vilenkin, 
Phys. Rev. Lett. {\bf 90}, (2003) 151301.}

\end{thebibliography}
\end{document}